\documentclass[useAMS,usenatbib]{mn2e}
\include{epsf}
\include{journals}
\usepackage{amsmath}
\usepackage{graphicx}
\usepackage{subfig}
\usepackage{float}
\usepackage{epstopdf}
\usepackage{bm}
\newcommand{\vect}[1]{\boldsymbol{\mathbf{#1}}}
\newcommand{\unit}[1]{\boldsymbol{\hat{\mathbf{#1}}}}
\newcommand{\mean}[1]{\boldsymbol{\bar{\mathbf{#1}}}}
\newcommand{\pec}[1]{\boldsymbol{\tilde{\mathbf{#1}}}}

\usepackage{xcolor}
\usepackage{makeidx}
\definecolor{myc}{rgb}{0.2,0.2,0.6}
\usepackage[ plainpages = false, pdfpagelabels, 
                 pdfpagelayout = OneColumn, 
                 bookmarks,
                 bookmarksopen = true,
                 bookmarksnumbered = true,
                 breaklinks = true,
                 linktocpage,
                 pagebackref,
                 colorlinks = true,
                 linkcolor = blue,
                 urlcolor  = blue,
                 citecolor = myc,
                 anchorcolor = green,
                 hyperindex = true,
                 hyperfigures
                 ]{hyperref} 

\newcommand{\bff}{\bf}
\renewcommand{\bff}{}   

\def\jref@jnl#1{{\rm#1}}

\def\aj{\jref@jnl{AJ}}                   
\def\araa{\jref@jnl{ARA\&A}}             
\def\apj{\jref@jnl{ApJ}}                 
\def\apjl{\jref@jnl{ApJ}}                
\def\apjs{\jref@jnl{ApJS}}               
\def\ao{\jref@jnl{Appl.~Opt.}}           
\def\apss{\jref@jnl{Ap\&SS}}             
\def\aap{\jref@jnl{A\&A}}                
\def\aapr{\jref@jnl{A\&A~Rev.}}          
\def\aaps{\jref@jnl{A\&AS}}              
\def\azh{\jref@jnl{AZh}}                 
\def\baas{\jref@jnl{BAAS}}               
\def\jrasc{\jref@jnl{JRASC}}             
\def\memras{\jref@jnl{MmRAS}}            
\def\mnras{\jref@jnl{MNRAS}}             
\def\pasa{\jref@jnl{PASA}}
\def\pra{\jref@jnl{Phys.~Rev.~A}}        
\def\prb{\jref@jnl{Phys.~Rev.~B}}        
\def\prc{\jref@jnl{Phys.~Rev.~C}}        
\def\prd{\jref@jnl{Phys.~Rev.~D}}        
\def\pre{\jref@jnl{Phys.~Rev.~E}}        
\def\prl{\jref@jnl{Phys.~Rev.~Lett.}}    
\def\pasp{\jref@jnl{PASP}}               
\def\pasj{\jref@jnl{PASJ}}               
\def\qjras{\jref@jnl{QJRAS}}             
\def\skytel{\jref@jnl{S\&T}}             
\def\solphys{\jref@jnl{Sol.~Phys.}}      
\def\sovast{\jref@jnl{Soviet~Ast.}}      
\def\ssr{\jref@jnl{Space~Sci.~Rev.}}     
\def\zap{\jref@jnl{ZAp}}                 
\def\nat{\jref@jnl{Nature}}              
\def\iaucirc{\jref@jnl{IAU~Circ.}}       
\def\aplett{\jref@jnl{Astrophys.~Lett.}} 
\def\apspr{\jref@jnl{Astrophys.~Space~Phys.~Res.}}
\def\bain{\jref@jnl{Bull.~Astron.~Inst.~Netherlands}} 
\def\fcp{\jref@jnl{Fund.~Cosmic~Phys.}}  
\def\gca{\jref@jnl{Geochim.~Cosmochim.~Acta}}   
\def\grl{\jref@jnl{Geophys.~Res.~Lett.}} 
\def\jcp{\jref@jnl{J.~Chem.~Phys.}}      
\def\jgr{\jref@jnl{J.~Geophys.~Res.}}    
\def\jqsrt{\jref@jnl{J.~Quant.~Spec.~Radiat.~Transf.}}
\def\memsai{\jref@jnl{Mem.~Soc.~Astron.~Italiana}}
\def\nphysa{\jref@jnl{Nucl.~Phys.~A}}   
\def\physrep{\jref@jnl{Phys.~Rep.}}   
\def\physscr{\jref@jnl{Phys.~Scr}}   
\def\planss{\jref@jnl{Planet.~Space~Sci.}}   
\def\procspie{\jref@jnl{Proc.~SPIE}}   

\title[Spiral driven peculiar velocity fields]{Spiral and bar driven peculiar velocities in Milky Way sized galaxy simulations}
\author[Grand et al.]{\parbox[t]{\textwidth}{
Robert J.J. Grand$^{12}$\thanks{robert.grand@h-its.org}, Jo Bovy$^{34}$\thanks{bovy@ias.edu}, Daisuke Kawata$^{5}\thanks{d.kawata@ucl.ac.uk}$, Jason A.S. Hunt$^5$, \\ Benoit Famaey$^6$, Arnaud Siebert$^6$, Giacomo Monari$^6$ and Mark Cropper$^5$} \vspace{10pt} \\
$^1$Heidelberger Institut f¡¯ur Theoretische Studien, Schloss-Wolfsbrunnenweg 35, 69118 Heidelberg, Germany\\
$^2$Zentrum f¡¯ur Astronomie der Universit¡¯at Heidelberg, Astronomisches Recheninstitut, M¡¯onchhofstr. 12-14, 69120 Heidelberg, Germany \\
$^3$Institute for Advanced Study, Einstein Drive, Princeton, NJ 08540, USA \\
$^4$John Bahcall Fellow \\
$^5$Mullard Space Science Laboratory, University College London, Holmbury St. Mary, Dorking, Surrey, RH5 6NT, United Kingdom \\
$^6$Observatoire astronomique de Strasbourg, Universit\'e de Strasbourg, CNRS, UMR 7550, 11 rue de l'Universit\'e, F-67000 Strasbourg, France}

\pagerange{\pageref{firstpage}--\pageref{lastpage}} \pubyear{2012}

\pagerange{\pageref{firstpage}--\pageref{lastpage}}\pubyear{2011}
\def\LaTeX{L\kern-.36em\raise.3ex\hbox{a}\kern-.15em
  T\kern-.1667em\lower.7ex\hbox{E}\kern-.125emX}
    
\begin{document}

\label{firstpage}
\maketitle

\begin{abstract}
We investigate the kinematic signatures induced by spiral and bar structure in a set of simulations of Milky Way-sized spiral disc galaxies. The set includes test particle simulations that follow a quasi-stationary density wave-like scenario with rigidly rotating spiral arms, and $N$-body simulations that host a bar and transient, co-rotating spiral arms. From a location similar to that of the Sun, we calculate the radial, tangential and line-of-sight peculiar velocity fields of a patch of the disc and quantify the fluctuations by computing the power spectrum from a two-dimensional Fourier transform. We find that the peculiar velocity power spectrum of the simulation with a bar and transient, co-rotating spiral arms fits very well to that of APOGEE red clump star data, while the quasi-stationary density wave spiral model without a bar does not. We determine that the power spectrum is sensitive to the number of spiral arms, spiral arm pitch angle and position with respect to the spiral arm. However, it is necessary to go beyond the line of sight velocity field in order to distinguish fully between the various spiral models with this method. We compute the power spectrum for different regions of the spiral discs, and discuss the application of this analysis technique to external galaxies.
\end{abstract}


\section{Introduction}

Understanding how large disc galaxies such as the Milky Way evolved into their present state is one of the principal aims of galactic astrophysics. In the Milky Way, particularly well known constraints for galaxy evolution include, {\bff for example}, the correlations of the age of stars with their kinematic and chemical properties, i.e., the age-velocity dispersion and age-metallicity relations. \citet{HNA09} showed from a combination of radial velocities derived from the metallicity data of F- and G-dwarf stars \citep{NMA04} and distances and proper motions from \emph{Hipparcos} \citep{VLF07}, that the velocity dispersion of stars increases with stellar age in all directions. Although the determinations of stellar age can be uncertain \citep[e.g.][]{RTT07,CSA11}, this positive trend between age and velocity dispersion is supported by other studies \citep[e.g.][]{DB98,SeG07,AB09}. Among others, a possible explanation for these trends is that a combination of spiral structure \citep{BW67,CS85} and giant molecular clouds \citep{SS51} effectively scatter stars in the {\bff planar} and vertical directions respectively, thus increasing the velocity dispersion of the stars while maintaining the observed radial to vertical velocity dispersion ratio. However, the mechanisms that drive these trends are still unknown.

Although the observed stellar kinematics are often described by a symmetric Gaussian in the radial and vertical directions, and a skewed Gaussian in the rotational direction, there is evidence that the velocity distribution in planar velocity space exhibits multiple streaming structures \citep[e.g.][]{DB98,CCB98,CCB99,FJL05,AFF08,BHR09,AMH15}, which have been attributed to the gravitational perturbations provided by non-axisymmetric structure. In particular, it has been reported that the local moving group known as the `Hercules' stream is likely caused by the constant periodic perturbations supplied by the outer Lindblad resonance of the bar \citep[e.g.,][]{D00,Fux01,MBS10}, which is thought to be close to the solar radius. The connection between the bar and such a moving group has been used to pin down some bar parameters, such as the pattern speed and orientation angle with respect to the solar position, {\bff either from the determination of the moving group position \citep[e.g.,][]{D00,B10,MAH13,AHD14} or the variation of the local Oort constants \citep{MQ07}.} 

Aside from the bar, spiral arms have been highlighted also to be the source of non-circular streaming motions \citep[see for example][]{DS04,BH10}, and in some cases has been directly linked to local moving groups such as the `Hyades' moving group \citep[e.g.,][]{S10a,PMF11}. In particular, spiral structure has been shown to change the guiding radii of disc stars by a process known as radial migration \citep{LBK72,SB02}. This occurs near the co-rotation radius of spiral arms, where the tangential gravitational force from spiral arms introduces a steady torque that imparts angular momentum changes on stars. Moreover, radial migration may be amplified if co-rotation resonances of spiral arms overlap other resonances such as the bar outer Lindblad resonance \citep[e.g.,][]{MF10}. Because radial migration occurs around regions of co-rotation, its effects depend on the location and number of resonances, and therefore the nature of the spiral arm itself. 

The most widely accepted theory of spiral structure is quasi-stationary density wave theory \citep{LS64,BL89}, in which the spiral arm is described as a stationary wave that rotates around the galaxy with a constant pattern speed without changing shape. However, {\bff there are conflicting results in the literature regarding the validity of density wave theory in external galaxies. For example,} application of the Tremaine-Weinberg method \citep{TW84} to external inclined galaxy discs typically favours radially decreasing spiral pattern speeds \citep[e.g.,][]{MRM06,SW11}. {\bff Furthermore, \citet{FR11} and \citet{FCK12} analysed the distribution of star forming tracers across spiral arms in external galaxies and found no evidence of the angular offsets between different star forming tracers \citep[although see][]{Eg09} predicted by theories of long-lived, rigidly rotating spiral arms \citep{R69}. From a numerical perspective, \citet{DB08} and \citet{DTP10} performed numerical simulations of isolated spiral galaxies and confirmed that tracer offsets are present in simulations with an imposed rigidly rotating spiral arm potential, like density waves, while they are absent in those without \citep[see also][]{GKC12}. However, some numerical studies of cosmologically simulated disc galaxies exhibit a distribution of young stars consistent with the predictions of classic density wave theory for long-lived spirals \citep{PGJ13}, and the velocity gradient found in the RAVE survey \citep{SFB11} has been shown to be reproducible with Lin-Shu density waves \citep{SFB12}. On the other hand, it is not clear that this is also the case for velocity fluctuations on larger scales, which we will consider in the present paper.}

In general, spiral arms in numerical simulations do not reproduce the long-lived {\bff single} density wave structure \citep[e.g.,][]{Se11}, and their scrutiny has spurred alternative ideas of the nature of spiral structure, such as multiple mode theory \citep{MFQD12,CQ12,S14}, manifold theory \citep{RG07,ARM09} and non-steady, co-rotating spiral arms \citep{WBS11,GKC11,GKC12,BSW12}. In the last description, spiral arms are transient, recurrent features that wind up and disappear with time, continually being replaced by new transient features that form from small perturbations amplified by a mechanism akin to swing amplification \citep{GLB65,JT66,T81,DO12}. These structures are found to be approximately co-rotating with the disc stars at every radius \citep[e.g.,][]{GKC11,GKC12}, and cause continual radial migration at every radius that manifests as large systematic streaming motions along the spiral arm \citep{GKC15,HKG15}, which is expected to be different from the peculiar motions induced by density wave spirals. The peculiar motions associated with different spiral arm models may therefore offer distinguishing observational predictions that can be tested particularly well in the Milky Way - the only galaxy for which we are able to get star by star information. {\bff The effects of different spiral arm models on the velocity distribution must be combined with those of spiral arm parameters such as the number of spiral arms, constraints on which have already been placed by analysis of peculiar motions of gas \citep[e.g.][]{PDA14} and stars in pencil beam surveys \citep[e.g.][]{MQ08} in numerical simulations. }

A useful way to characterise the peculiar motions in the disc was introduced by \citet{BBG14}, who studied the kinematics of red clump stars from APOGEE \citep{APM08}, abbreviated APOGEE-RC data, and computed the power spectrum of the peculiar line-of-sight (LOS) velocity field. {\bff From a comparison between the observed power spectrum and that of various disk galaxy models, they concluded} that the data favoured a bar only model that produced a peak peculiar velocity of $\sim 12$ $\rm km \; s^{-1}$ on scales of $\sim 2$ kpc. However, a successful model for the Milky Way ought to take into account the bar and spiral structure, both of which imprint their respective kinematical signatures in the peculiar velocity field. It is therefore key to model the perturbative effects of different spiral arm models both with and without a bar in order to understand the individual effects of each component. In this paper, we aim to build on the power spectrum analysis performed by \citet{BBG14} in order to compare the peculiar velocity fields produced in a range of spiral arm models with {\bff the APOGEE-RC LOS velocity data. {\bff In particular, we seek to establish whether or not spiral arms that form in $N$-body simulations are able to reproduce the characteristic features of the peculiar LOS velocity power spectrum.} In addition, we decompose the peculiar velocity field into radial and tangential components.} Scrutiny of these velocity fields allows us to quantify the induced peculiar motions in each model, in particular the $N$-body and spiral density wave models, and make observational predictions of each one {\bff that may be tested with $Gaia$ \citep{P01,LBB08} in the near future}. Furthermore, the simulations allow us to probe the effect of spiral arm pitch angle, spiral arm number and solar position with respect to the spiral arm on the resulting power spectrum, which may provide constraints on spiral arm parameters. We take further advantage of the simulations by computing the power spectra of both stars and gas over the whole disc (not just the solar neighbourhood region), which serve as observational predictions of the spiral-bar induced kinematics for external galaxies that may be tested by integral field spectroscopy (IFU) surveys such as MUSE \citep{BAA10} and CALIFA \citep{San12}. 

{\bff We note here that a limitation of this work is that we do not consider any external perturbations such as mergers and satellite interactions. However, satellite-disc crossings have been shown to generate, at a given position in the galaxy disc, ring-like features in planar velocity space \citep{MQW09,GMV12} and waves in vertical velocity \citep{WGY12,GMO13}. It is therefore possible that such interactions can contribute to the observed peculiar velocity features that we discuss in this paper. However, an investigation into this possibility is beyond the scope of this paper, and we leave this to future studies. }

\section{Simulations}
\label{secsim}

\begin{table*}
\centering
\caption{Table of simulation parameters. The columns are 1) Model name; 2) Model type; 3) Virial mass; 4) Total disc mass; 5) Radial scale length; 6) Spiral arm number; 7) Radial force ratio; and 8) Spiral arm pitch angle.}
\label{t1}
\begin{tabular}{c c c c c c c c}
\hline
Model & Type & $M_{vir}$ $(\rm \times 10^{12} M_{\odot})$ & $M_d$ $(\rm \times 10^{10} M_{\odot})$ & $R_d$ (kpc) & $m$ & $q$ & $\gamma$ \\
\hline
K14 & $N$-body, co-rotating spirals$+$bar & 2.5 & 5.0 & 2.5 & 2 & 0.16 & 10 \\  
F14 & Test particle, DW-like spirals & 0.6 & 5.13 & 2.0 & 2 & $0.23$ & 10 \\   
M15 & Test particle, DW-like spirals & 0.6 & 5.13 & 2.0 & 2 & $0.10$ & 10 \\  
G13F & $N$-body, co-rotating spirals & 1.5 & 5.0 & 3.5 & 4 & 0.08 & 19 \\   
G14 & $N$-body, co-rotating spirals  & 1.5 & 5.0 & 3.5 & 6 & 0.08 & 23 \\   
B15 & Test particle, bar only & \multicolumn{3}{|c|}{$V_c = 220\,\mathrm{km\,s}^{-1}$, flat rotation curve} & 2 & 0.015 & $-$ \\
\hline
Milky Way & $-$ & 1.0-2.0 & 6.0 & 2.15 & 2-4 & $-$ & 12 \\
\hline
\end{tabular}
\end{table*}

In this paper we analyse five simulations of isolated Milky Way-sized disc galaxies; three $N$-body {\bff simulations} and two test particle {\bff simulations}. The $N$-body simulations comprise a live disc component embedded in a static dark matter halo potential that follows the NFW profile \citep{NFW97}. The dark matter halo is modelled as a static potential in order to reduce computational costs and to avoid scattering of disc particles by dark matter particles, which are often restricted by computational limitations to be much more massive than disc particles. Further details of the simulation code are available in \citet{KG03,KOG13,KHG14}. 

The main simulation parameters relevant for this study are summarised in Table \ref{t1}. The fiducial simulation, labelled K14 \citep[see][for more details]{KHG14}, develops a bar and a prominent two-armed spiral structure. {\bff The spirals are transient features that wind up and disrupt owing to their co-rotation with the disc material. However, they are also recurrent features, and therefore the two-armed spiral morphology is sustained over the simulation duration.} Simulations G13F \citep[simulation F presented in][]{GKC13} and G14 \citep[presented in][]{GKC13b} develop transient spiral structure only, i.e., there is no bar. A further difference between these simulations is that simulation G13F has a central static stellar bulge component of mass $M_{b,\star} = 4.0 \times 10^{10}$ $\rm M_{\odot}$, and the dark matter distribution is slightly less centrally concentrated than in simulation G14. The difference in halo concentration means a lower disc-halo mass fraction at radii $R<10$ kpc in simulation G14 in comparison to G13F, which leads to a higher number of spiral arms in the former.

To determine the spiral arm strength in each model, we calculate the radial force per unit mass, $\mathbf{F}_R$, at a series of points distributed evenly in azimuth in a ring at a Galactocentric radius of 8 kpc, and then the ratio

\begin{equation}
q(R=8\; \rm{kpc}) = \bigg[ \frac{|\mathbf{F}_R(\theta) - \overline{\mathbf{F}}_R|_{max}}{\overline{\mathbf{F}}_R} \bigg]_{R=8\; \rm{kpc}}
\label{eqq}
\end{equation}
which describes the amplitude of radial force fluctuation caused by non-axisymmetric structure, $|\mathbf{F}_R(\theta) - \overline{\mathbf{F}}_R|_{max}$, as a fraction of the total axisymmetric radial force, $\overline{\mathbf{F}}_R$. We refer to $q$ as the radial force ratio, and take it to be the indicator of the strength of non-axisymmetric structure in each simulation.

\begin{figure}
\centering
\includegraphics[scale=0.28]{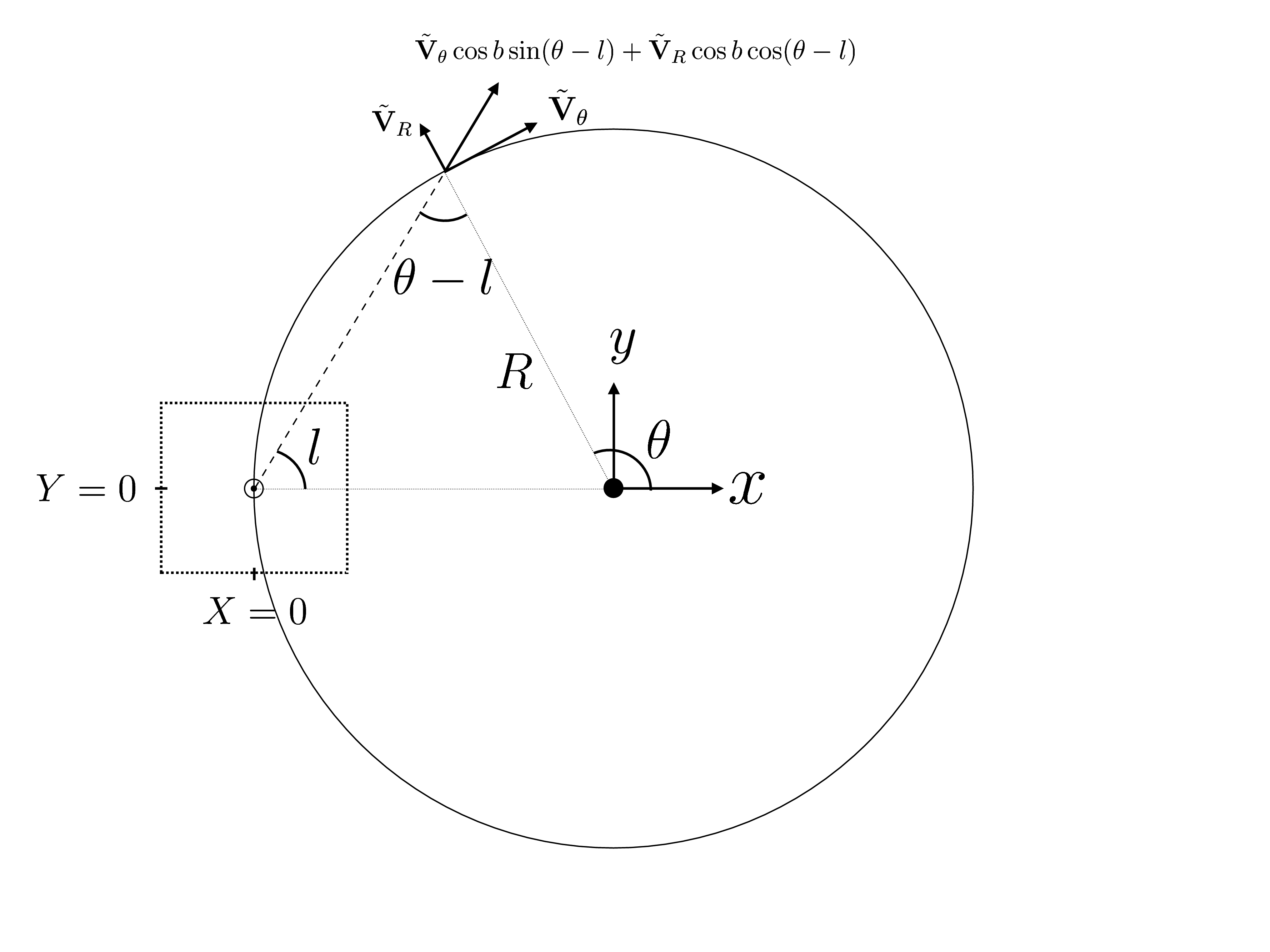} 
\caption{Two-dimensional schematic of the coordinates systems and peculiar velocity definitions described in the text. The Galactic centre and Solar position are marked by the solid circle and encircled dot, respectively. The line-of-sight to a star at solar radius in the disc plane and galactic longitude, $l$, is represented by the dashed line. {\bff The direction of rotation is clockwise.}}
\label{cs}
\end{figure}

Simulation F14, presented in \citet{FSF14}, is a test particle simulation in which the orbits of massless disc particles are integrated in a superposition of background axisymmetric (disc+bulge+dark matter halo) and spiral potentials. The spiral structure is that of a Lin-Shu type \citep{LS64}, {\bff with parameters chosen such that they match the RAVE results obtained by \citet{SFB12}, and it} rotates at a constant angular pattern speed that places the single co-rotation radius at $\sim 12$ kpc. {\bff However, we note that the radial velocity dispersion of the axisymmetric background model of F14 is lower than that of the Galaxy, and the radial force ratio is high. Therefore, we include another test particle simulation (Monari et al. in prep.), labelled M15, which has a radial force ratio equal to $0.10$ and a higher radial velocity dispersion. Note that for N-body simulations, the spiral arm gravitational potential does not follow a smooth cosine curve which is used to describe the spiral arm potential in the test particle simulations. Therefore, comparison between the $q$ values of $N$-body and test particle simulations may not be completely fair.}

{\bff The bar model from \citet{BBG14}, labelled B15, consists of a rotating quadrupole ($\propto \cos 2\phi$) with a pattern speed of $52.25\,\mathrm{km\,}^{-1}\,\mathrm{kpc}^{-1}$ and a radial force ratio of $1.5\%$, and makes an angle of $25^\circ$ with the Sun--Galactic-center line. This bar acts as a perturbation to a logarithmic axisymmetric potential with a circular velocity of $220\,\mathrm{km\,s}^{-1}$. The bar's effect on the kinematics of a population of stars with a radial velocity dispersion of $31.4\,\mathrm{km\,s}^{-1}$ is computed using \textsc{galpy}\footnote{http://github.com/jobovy/galpy~.} \citep{B15}.}

\section{The Peculiar Velocity Power Spectrum}

In this section, we outline the procedure for calculating the peculiar velocity power spectrum of the simulation data. The procedure consists of two steps: 1) to calculate a peculiar velocity field of a given region of a simulated galaxy, and 2) to apply a two dimensional Fourier transform to the peculiar velocity field, which is converted to a one dimensional power spectrum. Below we give a brief description of the implementation of this procedure \citep[see][for more details]{BBG14}.

\begin{figure*}
\centering
\includegraphics[scale=1.4]{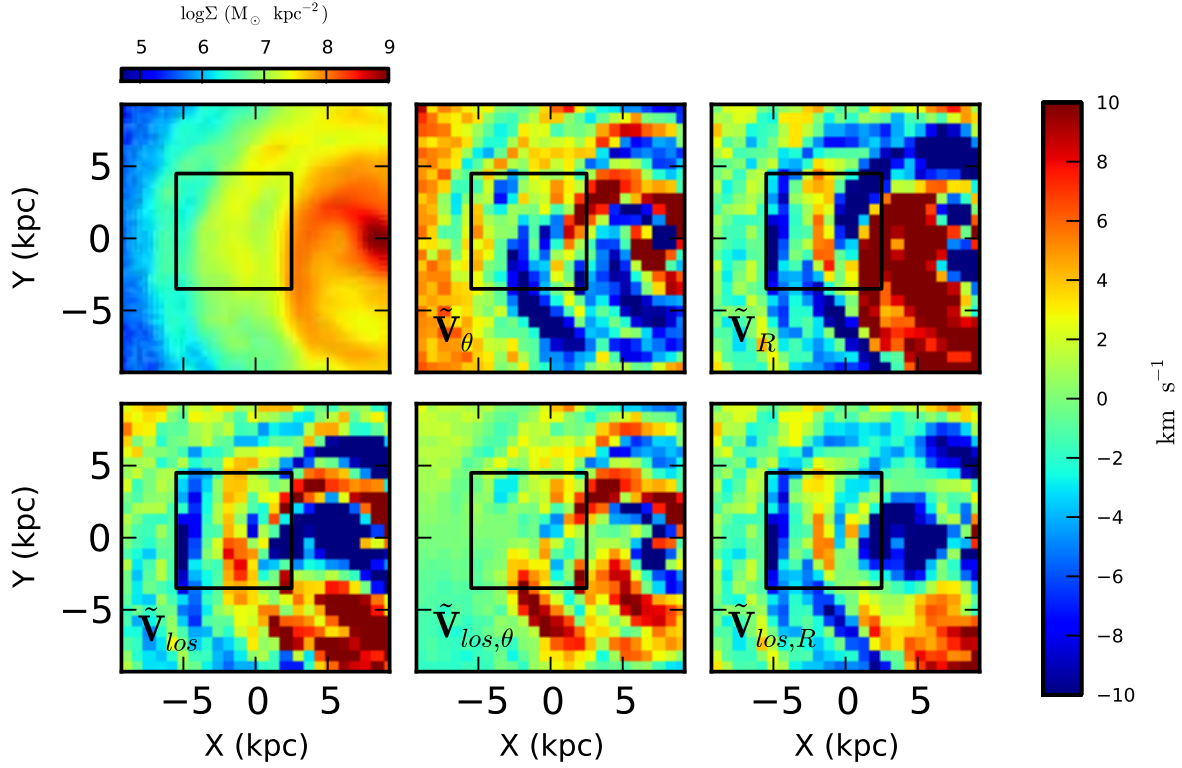} \vspace{-10.0mm}
\caption{Simulation K14: {\bff log {\bff projected surface density, $\Sigma$} (top-left), the magnitude of the intrinsic rotational and radial peculiar velocity fields (top-middle and -right panels respectively), the peculiar LOS velocity field (bottom-left), the rotational and radial component of the peculiar LOS velocity field  (bottom-middle and -right panels respectively). }}
\label{vf1}
\end{figure*}

\begin{figure*}
\centering
\includegraphics[scale=1.4]{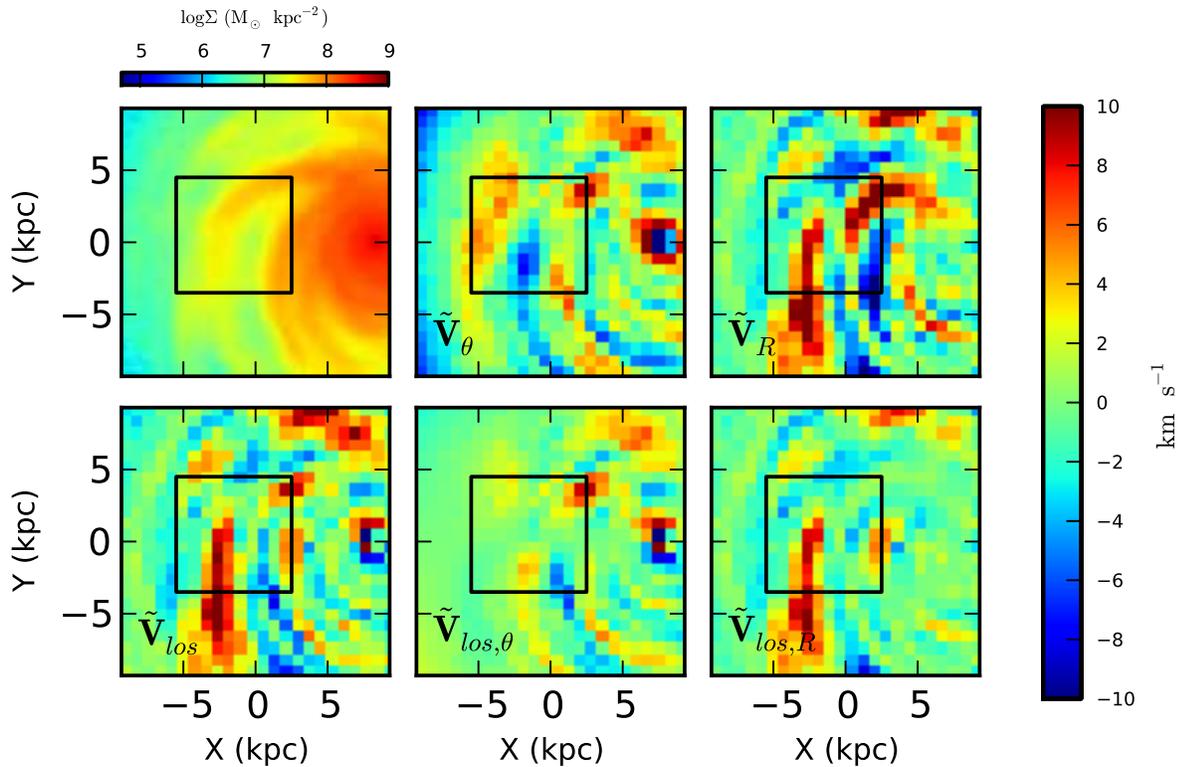} \vspace{-10.0mm}
\caption{As Fig. \ref{vf1}, but for the spiral analysed in simulation G13F.}
\label{vf2}
\end{figure*}

\subsection{The Peculiar Velocity Field}
\label{secpvf}

The total peculiar velocity of a star is defined as the deviation from the mean velocity in the radial, tangential and vertical directions, $\unit{R}$, $\unit{\theta}$ and $\unit{Z}$, respectively. In a rotationally supported axisymmetric disc, the mean velocities are $\mean{V}_R = \mean{V}_Z = 0$, and $\mean{V}_{\theta}$ is non-zero and depends on the circular velocity and the asymmetric drift. We therefore define the peculiar velocities of the $i$-th particle to be

\begin{equation}
\begin{aligned}
\pec{V}_{R,i} = \vect{V}_{R,i},  \\
\pec{V}_{{\theta},i} = \vect{V}_{{\theta},i} - \mean{V}_{\theta},  \\
\pec{V}_{Z,i} = \vect{V}_{Z,i}. 
\end{aligned}
\label{eqp}
\end{equation}
In our coordinate system (with clockwise rotation), {\bff positive velocities indicate motion away from the Galactic centre in the radial direction, motion faster than the mean rotational velocity in the tangential direction, and motion in the positive $Z$ direction.} {\bff To calculate the LOS components of the {\bff intrinsic (or galacto-centric)} peculiar velocities defined in equation (\ref{eqp}), we define an angle, $\theta$, to be the angle between the line that connects the star to the galactic centre and the line defined by $y=0$ for positive $x$, of a cartesian grid whose origin is at the galactic centre (see Fig. \ref{cs} for a diagram illustrating the coordinate system). The peculiar LOS velocity can be decomposed into a radial and rotational component given by }

\begin{subequations}
\begin{align}
\pec{V}_{los,R} = \pec{V}_{R} \cos b \cos (\theta - l) , \\
\pec{V}_{los,\theta} = \pec{V}_{\theta} \cos b \sin (\theta - l) , 
\end{align}
\end{subequations}
{\bff where $l$ and $b$ are the Galactic longitude and Galactic latitude coordinates, respectively.} Once the radial and rotational LOS peculiar velocities are computed for the star particles, the particles are binned onto a grid of spatial resolution equal to $0.8$ kpc and the mean LOS velocity value is taken for each bin.

\subsection{Computing the Power Spectrum}
\label{seccps}

Once we have calculated the peculiar velocity field, $f_{pq}$, on a grid of $N \times N$ points, we calculate the two dimensional Fourier transform, $F_{lm}$, as

\begin{equation}
F_{lm} = \sum _{p=0}^{2N} \sum _{q=0}^{2N} f_{pq} e^{\pi i [p k_x + q k_y] / k_{xy}^{max}}, 
\end{equation} 
where $(k_x, k_y) = (k_x^{max} l/N, k_y^{max} m/N)$. Here, $k_{x,y}^{max}$ is defined to be the maximum frequency value that can be sampled (the Nyquist frequency). The grid has been padded at the end by $N$ zeros in both directions, in order to remove the signal wrap-around pollution caused by the assumption of the convolution theorem that the signal is periodic. The Fourier-transformed velocity field forms a separate $N \times N$ grid that stretches from $k_{xy} = 0$ to $k_{xy}^{max}$ on both the $x$ and $y$ axis, with spacings of $\Delta k = k_{xy}^{max} / N$. For each value $F_{lm} = F(k_x,k_y)$, the power may be estimated as

\begin{equation}
P(k_x,k_y) = (4 \pi)^2 \cdot |F(k_x,k_y)|^2.
\end{equation}
The one dimensional power spectrum, $P(k)$, is estimated by averaging the power of the two dimensional power spectrum in annuli of $k = \sqrt{k_x^2+k_y^2}$ \citep[see][for more details]{BBG14}.

\subsection{The Choice of Solar position}

We apply the methodology described in Sections \ref{secpvf} and \ref{seccps} to snapshots of the simulations described in Section \ref{secsim}. The patch of disc in which we choose to apply the Fourier analysis extends from $-5.5$ to $2.5$ kpc in $X$ and $-3.5$ to $4.5$ kpc in $Y$, where $X$ and $Y$ represent {\bff heliocentric cartesian coordinates {\bff (as opposed to the galactocentric coordinates, $(x,y)$, referred to in the previous section)}, i.e., $(X_{\odot},Y_{\odot}) = (0.0,0.0)$. From the Solar position, positive $X$ points in the direction of the Galactic centre, and positive $Y$ points in the direction of rotation {\bff (see the dashed box in Fig. \ref{cs})}. The patch provides} adequate coverage of the spiral arm as viewed from the solar position and is similar in size to the disc patch analysed by \citet{BBG14} and the {\bff volume covered by the APOGEE-RC data}. The bin size of the disc patch is $0.8$ kpc, and we include all stars within $|z| < 250$ pc {\bff as} in \citet{BBG14}. 

We set the position of the Sun at the galactocentric radius of 8 kpc. The azimuthal angle of the Solar position is determined so that there is a spiral arm at a distance of 4 kpc from the Sun in the direction $l=90$ degrees \citep[similar to that found by][who fit the spiral arm using parallaxes of high mass star forming maser sources]{RMB14}. 

\begin{figure*}
\centering
\includegraphics[scale=1.4]{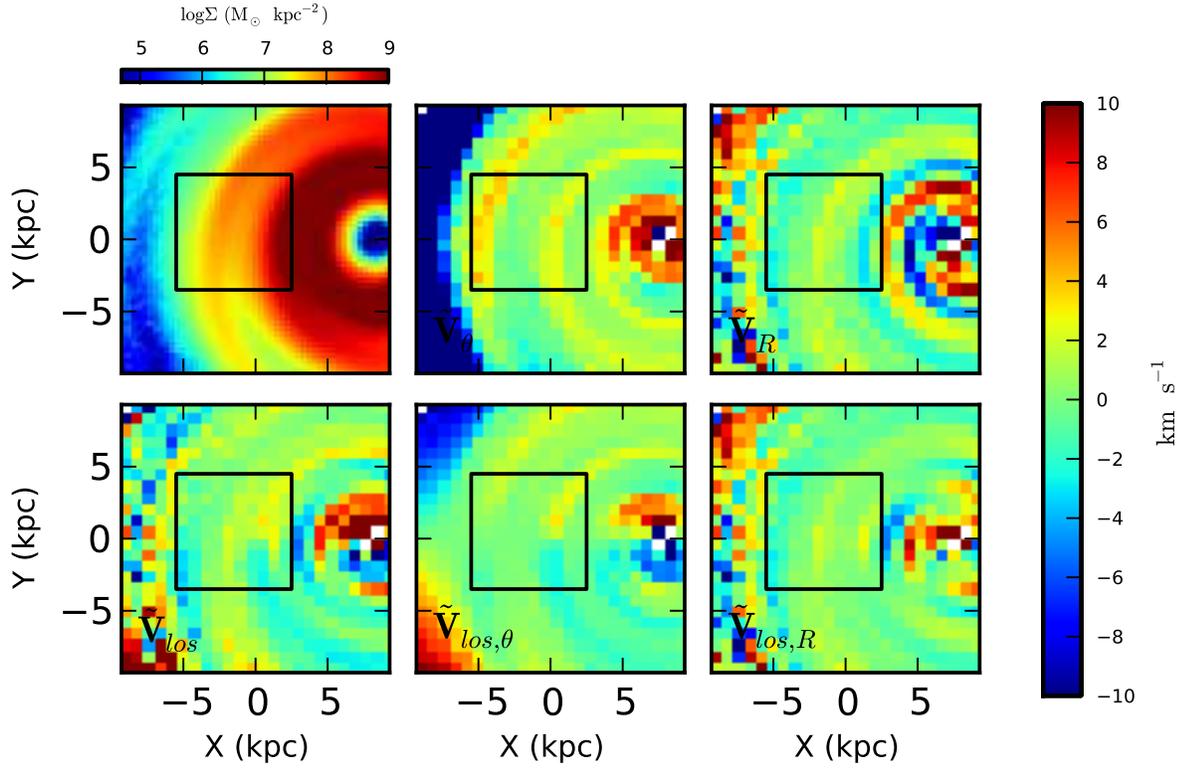} \vspace{-10.0mm}
\caption{As Fig. \ref{vf1}, but for the spiral analysed in simulation F14.}
\label{vf3}
\end{figure*}

\begin{figure*}
\centering
\includegraphics[scale=1.4]{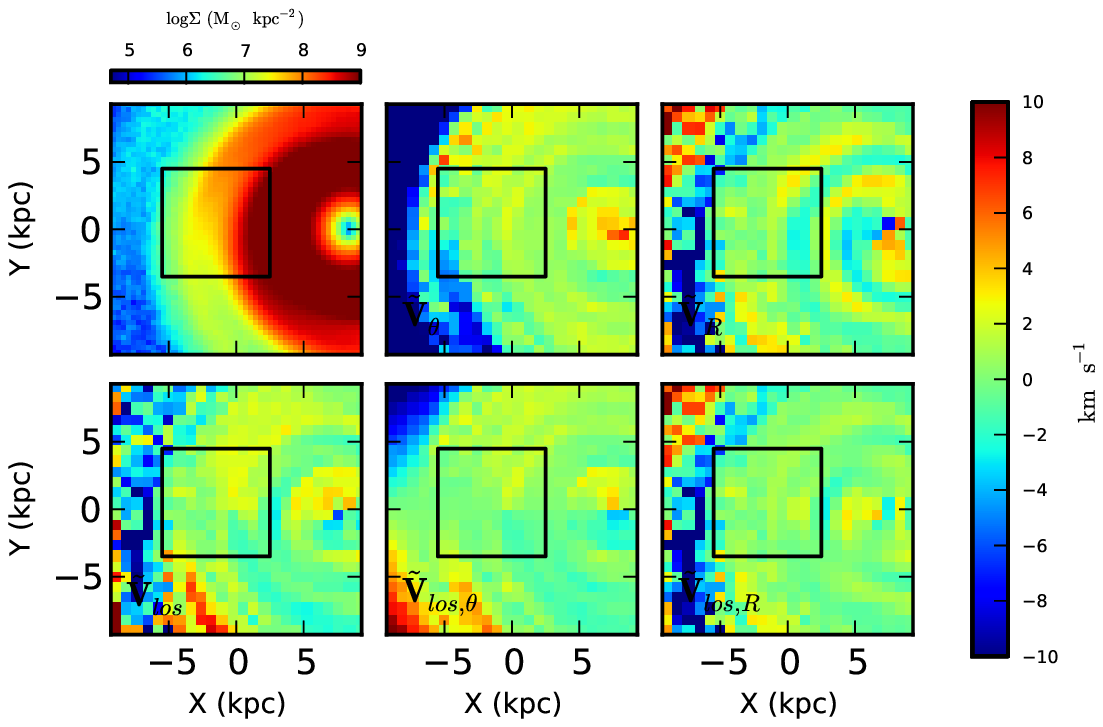} \vspace{-10.0mm}
\caption{As Fig. \ref{vf1}, but for the spiral analysed in simulation M15.}
\label{vf5}
\end{figure*}

\begin{figure*}
\centering
\includegraphics[scale=1.4]{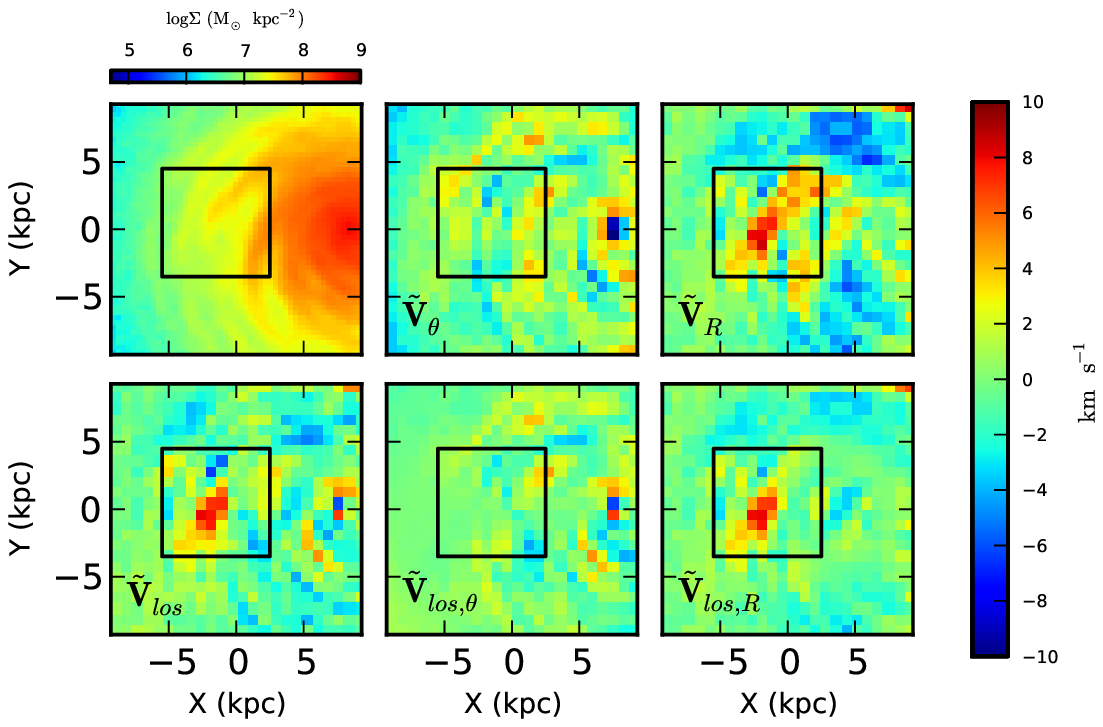} \vspace{-10.0mm}
\caption{As Fig. \ref{vf1}, but for the spiral analysed in simulation G14.}
\label{vf4}
\end{figure*}

{\bff We analysed many snapshots in each simulation, and found that the power spectrum varies between snapshot for the $N$-body simulations (see Section \ref{specdep}). However, the goal of this paper is to determine whether a particular model can reproduce the features of the observed power spectrum. Therefore, for each simulation we focus on the snapshot which most closely matches the observational data. Contrary to the $N$-body simulations, it is not necessary to explore many snapshots of the test particle simulations because the velocity field does not change with time after the initial instability caused by the introduction of the rigid spiral potential disappears. Therefore, we show the results at a snapshot after the system is well relaxed.}
 
\section{Results}

\begin{figure*}
\centering \hspace{15.0mm}
\includegraphics[scale=1.7]{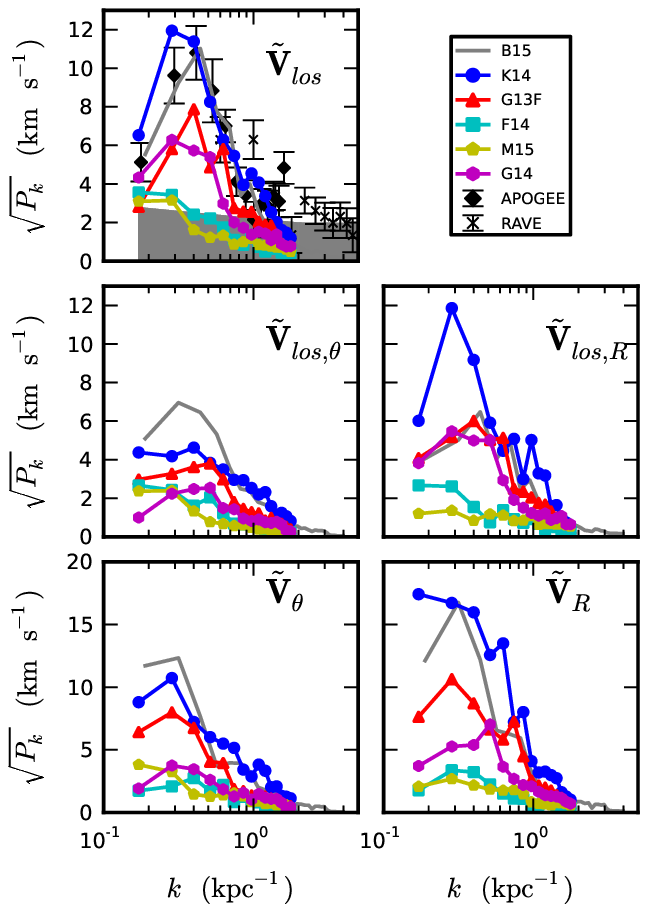} \vspace{-20.0mm}
\caption{The azimuthally averaged one-dimensional power spectrum of: the peculiar LOS velocity field (top-left), the rotational and radial components of the peculiar LOS velocity field (middle-left and -right) and the intrinsic rotational and radial peculiar velocities (bottom-left and -right). Each solid curve denotes the power spectrum of the respective field from a simulation model. {\bff The observed power spectrum measured by \citet{BBG14} from APOGEE-RC and RAVE data is shown in black points with associated error bars. The shaded area marks the $95\%$ noise region of the data.}}
\label{ps1}
\end{figure*}

Fig. \ref{vf1} shows the peculiar velocity fields and particle distribution of a snapshot in simulation K14. The plots are centred on our chosen Solar position (which is placed at 8 kpc from the Galactic centre), and the black box highlights the patch of the disc for which we calculate the power spectrum. The {\bff top-left} panel of Fig. \ref{vf1} shows {\bff the log particle surface density, $\Sigma$}. The {\bff top-middle and -right} panels of Fig. \ref{vf1} show the peculiar rotational and radial velocity respectively, which reveals that the spiral structure (as well as the bar) are associated with clear features in both velocity fields. For example, strong negative radial velocities (towards the galactic centre) trace the leading edge of the {\bff Perseus-like spiral arm located at a distance of about 4 kpc from the solar position in the direction $l=90$, similar in location to the Perseus spiral arm in the Milky Way \citep{RMB14},} whereas the velocity field changes to positive radial velocities (towards the galactic anti-centre) on the trailing edge of the spiral arm \citep{GKC13b}. 

The {\bff bottom-middle and -right panels of Fig. \ref{vf1} show the radial and rotational components of the peculiar LOS velocities. Similar to the intrinsic peculiar velocity fields, the peculiar LOS velocity fields show significant fluctuations,} although the fluctuations are significantly reduced in regions where the LOS velocity is not sensitive to the rotational and radial velocity, i.e., along the $X-$axis for $Y=0$ and an arc region whose galactocentric radius is similar to that of the Sun, respectively. The {\bff bottom-left} panel of Fig. \ref{vf1} shows the total LOS peculiar velocity that results from the summation of the {\bff rotational and radial LOS components}. {\bff Figs. \ref{vf2}, \ref{vf3}, \ref{vf5} and \ref{vf4} show the same velocity fields as Fig. \ref{vf1} for the spirals examined in G13F, F14, M14 and G14, respectively. {\bff All simulations show velocity fields that trace the spiral structure. Similar to K14, the radial and rotational components of the intrinsic peculiar velocity fields of G13F (Fig. \ref{vf2}) and G14 (Fig. \ref{vf4}) exhibit radial velocities that generally point outward (inward) on the trailing (leading) side of spiral arms, and rotational velocities which are slower (faster) than the mean on the trailing (leading) side of spiral arms \citep{GKC13b}. These trends of intrinsic peculiar velocities is evident in the LOS velocities as well. The main difference that can be seen by eye between Figs. \ref{vf1} and \ref{vf2} is the lack of velocity structure in the central region of the galaxy (where $4 < X < 8$ and $-4 < Y < 4$) owing to the presence (lack of) a bar in K14 (G13F). In comparison to the $N$-body simulations, the test particle simulations F14 (Fig. \ref{vf3}) and M15 (Fig. \ref{vf5}) show very clear spiral patterns in the surface density and velocity fields, which arises from the analytically defined spiral density perturbation. In these models, the mean radial velocity inside the spiral arms generally points inward (negative velocity), whereas it points outward (positive velocities) in the inter arm regions \citep{SFB12}. The main difference between the peculiar velocity fields of F14 and M15 is that the magnitude of the fluctuations is larger for F14, in particular in the rotational component, which is a result of the larger spiral amplitude of F14 relative to M15. }   

Fig. \ref{ps1} shows the azimuthally averaged one-dimensional power spectra of the velocity fields for K14 shown in Fig. \ref{vf1}, in addition to the {\bff power spectra of the velocity fields in a similar region} of the other simulations. The top-left panel plots power spectra of the simulations alongside the best fitting bar model of \citet{BBG14}, B15, and spectra calculated from the observed peculiar velocities from the APOGEE and RAVE surveys \citep{BBG14}. {\bff The power spectrum of the fiducial simulation K14 exhibits a turnover at peak power in the range $\sim 0.3$ - $0.4$ kpc$^{-1}$, and is a good match to the data and bar model of B15, although the power at scales of $k\sim 0.3$ kpc$^{-1}$ appears to be larger than that of the data by about $2$ km s$^{-1}$.} We have confirmed that the power spectrum does not change significantly for {\bff the case of a more limited volume like the region covered by the APOGEE-RC data.}

Interestingly, {\bff the power spectrum of the peculiar LOS velocity field in the spiral-only N-body simulation, G13F, exhibits a similar peak shape to} the observed power spectrum, { \bff although the amplitude on all scales is smaller than that of K14. This reflects the strength of non-axisymmetric structures, which is larger for K14 than for G13F (see the q column of Table. 1).} In contrast to the $N$-body simulations, the {\bff power spectra of the} density wave models of F14 and M15 {\bff show increasing power on progressively larger scales. This indicates that the density wave-like spiral models do not fit the data well, which is consistent with other density wave - test particle models shown in \citet{BBG14}.}

\begin{figure}
\centering
\includegraphics[scale=1.1]{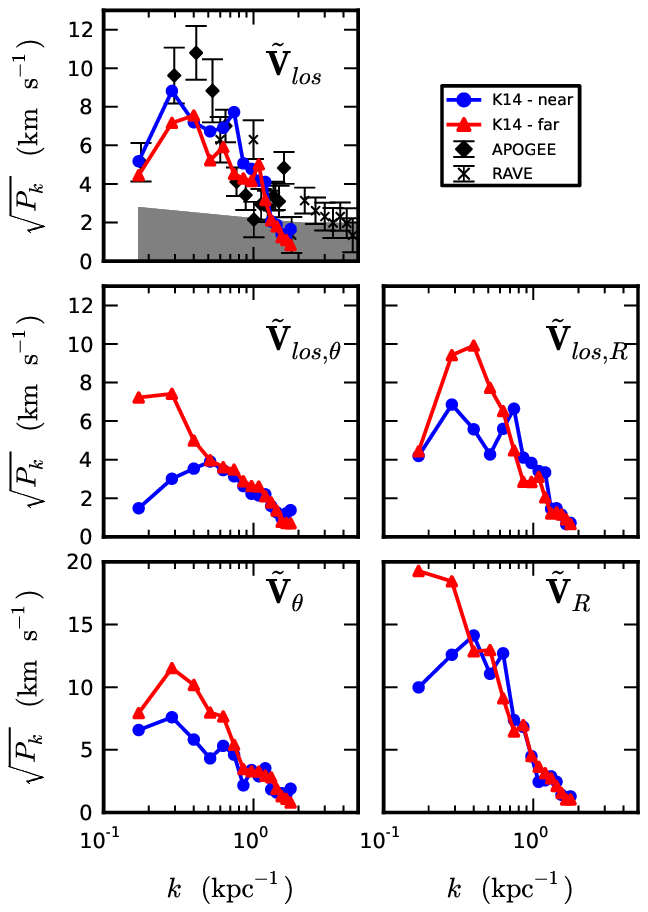} \vspace{-20.0mm}
\caption{{\bff The same as Fig. \ref{ps1} but comparing the power spectra of the fiducial spiral arm of K14 from two different solar positions: a position nearer to (blue) and a farther from (red) the spiral in comparison to the fiducial K14 position shown in Fig. \ref{ps1}.}}
\label{ps2}
\end{figure}

\begin{figure}
\centering
\includegraphics[scale=1.1]{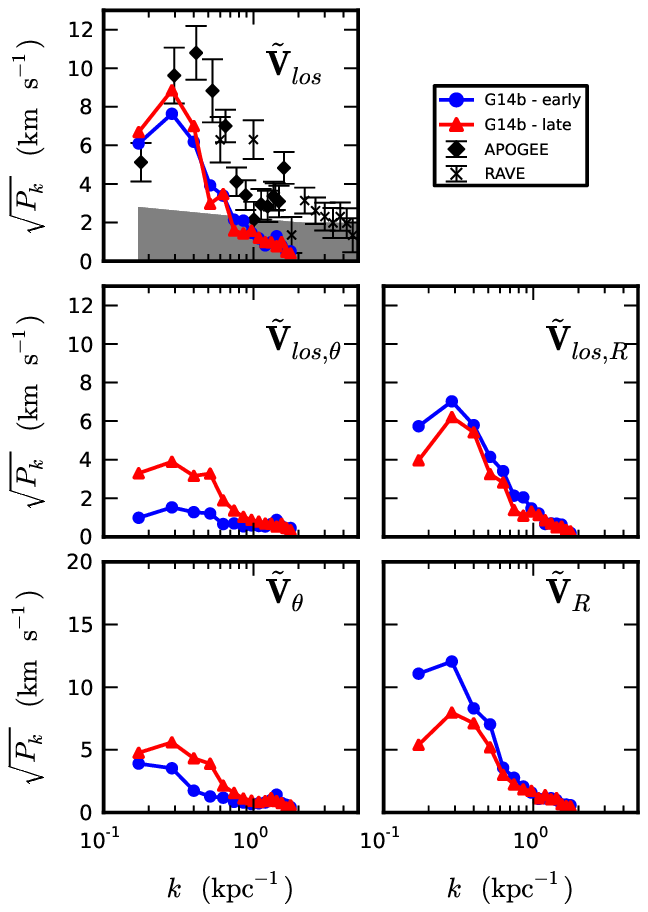} \vspace{-20.0mm}
\caption{The same as Fig. \ref{ps1} but comparing the power spectrum {\bff of spiral arm G14b. The blue corresponds to the spiral arm at the earlier stages of formation when the pitch angle is higher, whereas the red line corresponds to the same spiral arm at a later stage of formation when the pitch angle is lower}.}
\label{ps3}
\end{figure}

The above results show that in addition to the bar only model, B15, the barred-spiral $N$-body simulation, K14, is able to reproduce the observed peculiar LOS power spectrum very well, in contrast to the cosmological $N$-body simulation \citep{GC11} investigated by \citet[][]{BBG14}. This suggests that the observed peculiar motions {\bff can be explained by either a combination of a bar and transient, co-rotating spiral arms \citep{GKC12,KHG14} or the bar component only as found by \citep[][B15]{BBG14}.} In contrast, the observed power spectrum cannot be explained by the stellar motion induced by the density wave-like spiral arms in test particle simulations F14 and M15. On the other hand, the power spectrum of simulation G13F (and to an extent G14) shows evidence of a turnover in power on scales of $k \sim 0.4 \; \rm{km\;s^{-1}}$, which is the characteristic feature of the observed power spectrum. Therefore, it is possible that the main features of the observed power spectrum can be explained purely by the systematic stellar motion induced by co-rotating spiral arms seen in simulations \citep[e.g.,][]{GKC11,GKC12,GKC13b}. We note that this study is limited to our current set of simulations, and we have not extensively searched for a $N$-body simulation model with transient spiral arms only that can better reproduce the observed power spectrum. By the same token, we have analysed only two test particle simulations with density wave-like spiral arms, and therefore cannot state outright that no density wave configuration can explain the observed power spectrum. Our study highlights that the observed power spectrum of the APOGEE-RC data can be explained not only by the bar induced velocity fields, but also by the velocity field induced by the spiral arms. It is therefore interesting to explore many more different models of spiral arms and compare them with the observations.

The middle (bottom) panels of Fig. \ref{ps1} {\bff show the power spectra of the} radial and tangential components of the LOS (intrinsic) peculiar velocity fields. {\bff The power spectra of model F14 and M15 show much less power in the intrinsic velocity fields than the simulation K14 and G13F, which results in less power in the LOS rotational and radial velocities, the latter of which seems to drive the shape of the power spectrum in the top-left panel. }

{\bff It is interesting to note that the power in K14 is dominated by the radial LOS velocity, despite the relatively strong intrinsic rotation peculiar velocity.} This severe reduction in power between intrinsic and LOS rotational velocity may be explained by the `flip' in LOS velocity direction that occurs either side of $Y=0$ in the {\bff bottom-middle} panel of Fig. \ref{vf1}. For example, the positive intrinsic rotation velocity region at $(X,Y)=(2,4)$ in the {\bff top-middle} panel of Fig. \ref{vf1} corresponds to a receding (positive) LOS velocity in the {\bff bottom-middle} panel of Fig. \ref{vf1}. However, the strong negative intrinsic rotation velocity region at $(X,Y)=(-2,-2)$ in the {\bff top-middle} panel of Fig. \ref{vf1} indicates a slow rotational peculiar velocity, which leads to a receding (positive) LOS velocity as viewed from the solar position. {\bff It appears that this `flip' in velocity direction caused by the conversion from intrinsic to LOS velocity can affect the resultant power spectrum, which therefore is dependent on the selected area of analysis and does not reflect the true effects of non-axisymmetric structure on the intrinsic radial and rotational velocity fields. It is therefore important to observe the 3D velocity field in a larger area, which will soon be possible for the observational data provided by $Gaia$.}

\subsection{Other Power Spectrum dependencies}
\label{specdep}

\subsubsection{Spiral arm number}

Fig. \ref{ps1} contains the power spectra of simulation G14, which is a flocculent $m=6$ spiral galaxy simulation suitable for comparison with G13F and K14 in order to study the effect of spiral arm number on the power spectrum. The most pronounced difference is visible in the bottom-right panel of Fig. \ref{ps1}, which shows that the power in intrinsic peculiar radial velocity peaks at smaller scales, $k =0.5$ kpc$^{-1}$, than G13F and K14, which peak at $k =0.3$ - $0.4$ {\bff kpc$^{-1}$}. This indicates that flocculent spiral structure produces more power on small scales owing to the smaller inter arm separation, which limits the spatial extent of the peculiar velocity fluctuations. However, the power spectrum of the total LOS peculiar velocity is not distinguishable in shape from simulations K14 and G13F.

\subsubsection{Proximity of Spiral arm}

Fig. \ref{ps2} shows how spiral arm proximity with respect to the solar position affects the power spectrum in simulation K14. We consider two configurations in addition to the fiducial power spectrum: one in which the disc is rotated 20 degrees clockwise from the fiducial setup, and one in which is rotated 20 degrees anti-clockwise from the fiducial setup. In this way we obtain three solar positions with respect to the spiral arm: a {\bff far}, an intermediate (fiducial) and a {\bff near position}. It is clear from Fig. \ref{ps2} that neither the near nor far configuration reproduces the observed power spectrum; the far configuration, though it reveals similar intrinsic peculiar velocity power spectra, does not reach the height of the observed peak of the LOS peculiar velocity spectrum, whereas the near configuration exhibits a contrasting power spectrum in all peculiar velocity fields. Of particular note is the double peak structure in radial and total LOS peculiar velocity, which may arise from the faint stellar filamentary structure on the leading side of the main `Perseus'-like spiral. Such a structure in close proximity of the main spiral may produce similar effects to the flocculent spiral structure described above.

\subsubsection{{\bff Evolutionary stage of the Spiral arm}}

As highlighted by \citet{BSW12} and \citet{GKC13}, spiral arms in $N$-body simulations continuously wind up from the moment that they begin to form until they are disrupted, and therefore each spiral arm sweeps though a range of pitch angles during its lifetime. In this section, we explore the dependency of the power spectrum on different spiral evolutionary stages (and therefore, pitch angle), which has been previously reported to affect the velocity distribution of the solar neighbourhood \citep{DS04}. We focus on a spiral arm from simulation G14 whose winding evolution is studied in detail by \citet{GKC13b}. Note that this spiral arm is {\bff a different spiral arm from the spiral studied above (labelled G14), therefore we denote this spiral arm G14b to distinguish it from the original. Spiral G14b} is in a less flocculent region of the disc\footnote{Although simulation G14 \citep{GKC13b} is reported to have $m=6$ spiral arms, this number is not always constant with time nor is the spiral structure perfectly symmetric.}, which facilitates the study of the effect of pitch angle. 

Fig. \ref{ps3} shows the power spectra of the same size velocity field patch at two different evolutionary stages of the same spiral arm: an early stage during which the spiral has a relatively high pitch angle, and a later stage when the spiral arm is relatively tightly wound. Both the early and late stage spiral arms produce very similar LOS velocity power spectra. However the intrinsic radial and rotational velocity fields reveal that the spiral of larger pitch angle (G14b-early) produces more power in the radial peculiar velocity field than the spiral of lower pitch angle at the later spiral evolutionary stage (G14b-late). It seems plausible that there exist spiral-driven time-dependent peculiar velocity fields, which evolve with the winding spiral. Indeed, Fig. \ref{rth} demonstrates a scenario in which the open spiral arm (G14b-early) drives strong systematic radial motions over a large radial range, which is reflected in the bottom-right panel of Fig. \ref{ps3}. As the spiral arm winds to lower pitch angles (G14b-late), the velocity vectors appear to re-orient themselves to a more tangential direction, which has the effect of reducing the power in the radial peculiar velocity spectrum while increasing the power in the rotational peculiar velocity spectrum (though the difference shown in the bottom-left panel of Fig. \ref{ps3} is slight). { \bff The leading cause of this difference may be related to either the different stages of spiral arm formation and disruption \citep[see, for example,][]{BSW12}, or the different pitch angle of the spiral arm at these instances. A further detailed study is required to determine the cause.}

Although the above analysis indicates that the peculiar velocity field is dependent on spiral arm number, solar position and spiral arm pitch angle, we caution that further work covering a much larger suite of simulations is required to fully work out the dependencies and degeneracies of the peculiar velocity power spectrum. On the other hand, this means that the stage of spiral arm evolution may be able to be constrained by analysis of the peculiar velocity fields presented in this paper.

\subsection{External galaxies}

\begin{figure}
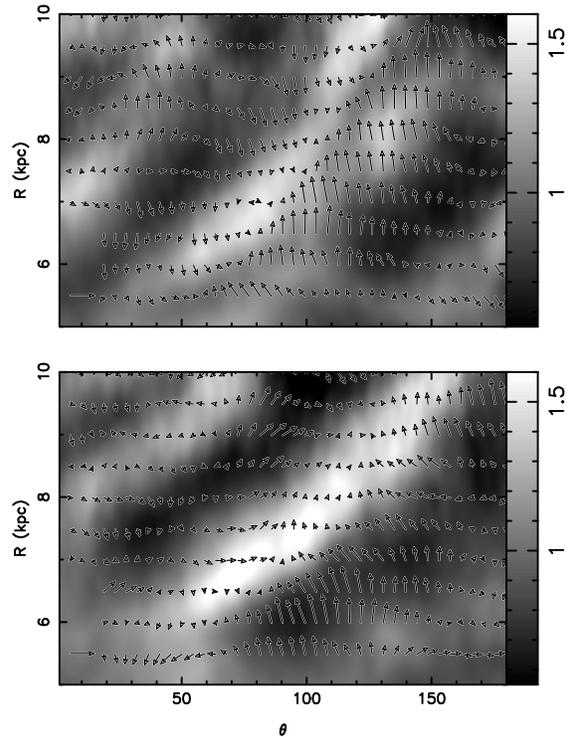

\centering 
\subfloat{\includegraphics[scale=0.4]{figures/rthst002698}}\\\vspace{-12.0mm}
\subfloat{\includegraphics[scale=0.4]{figures/rthst002753}}\\
\caption{Close up density map in polar coordinates of the spiral arm in simulation {\bff snapshots G14e (upper) and G14l (lower)}. The top panel shows the forming spiral arm in front of the solar position at $(R,\theta) \sim (8, 150)$. The bottom panel shows the same spiral arm at a later time when it is fully formed. Over-plotted is the mean peculiar velocity field, which is shown to change orientation as the spiral arm evolves.}
\label{rth}
\end{figure}

\begin{figure}
\centering
\subfloat{\includegraphics[scale=1.25]{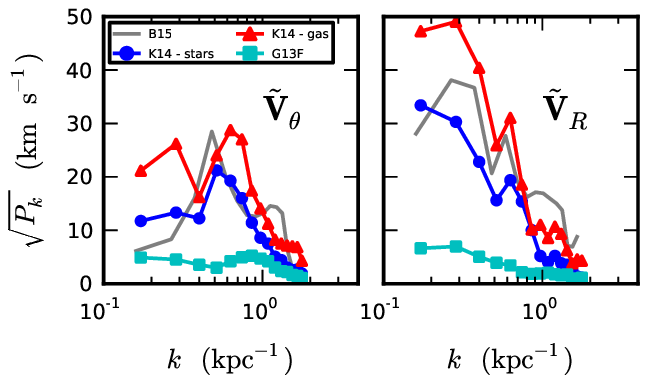}} \\
\subfloat{\includegraphics[scale=1.25]{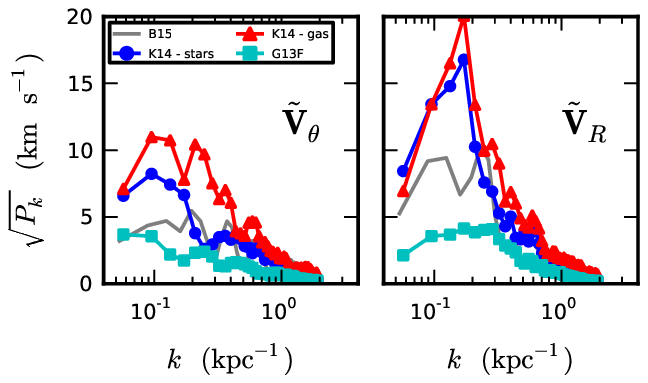}}\\
\subfloat{\includegraphics[scale=1.25]{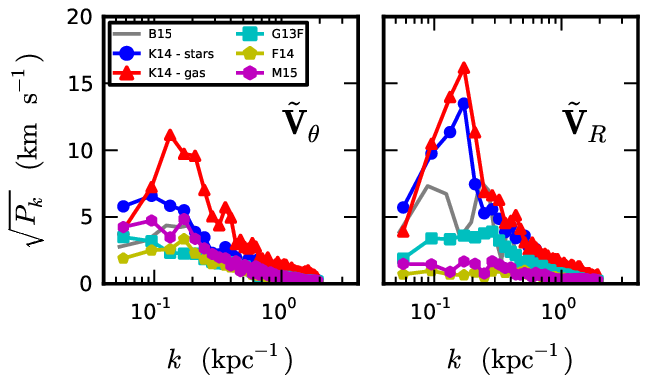}}\\
\caption{The power spectrum of the intrinsic peculiar rotational and radial velocity fields of the central region of the disc (top), the whole disc (middle) and the whole disc with the central region cut out (bottom). Note the change in the x-axis and y-axis ranges.}
\label{ps5}
\end{figure}

The power spectrum analysis presented so far is a useful tool for quantifying the perturbed velocity field in the Milky Way, in which we can observe only a local patch of the disc at present \citep[see also][]{BBG14}. For external galaxies, however, we have available the whole disc region on which to apply the power spectrum analysis. Therefore different regions of the disc, such as the centre, can be examined as well as the whole disc. {\bff The power spectrum analysis of a region that extends over the whole disc is more robust than that of the local disc patch analysed in the previous sections, which can be sensitive to the choice of patch (see Fig. \ref{ps2}).} In Fig. \ref{ps5}, we show the power spectra of the rotational and radial peculiar velocity fields {\bff for different regions of the disc. In the top panel of Fig. \ref{ps5}, we present the power spectra of the central $8\times 8$ kpc region of the disc, i.e, extending from $4.0$ to $12.0$ kpc in $X$ and from $-4.0$ to $4.0$ kpc in $Y$, for models K14 and G13F (we do not show the power spectra of the density wave models, F14 and M15, because these models follow particle motion at $R>4$ kpc only). For K14 we analyse also the power spectrum of the gas velocity field.} There are two main characteristics of the peculiar velocity power spectrum of the central region: the first is the large power in the fluctuations of both rotational and radial peculiar velocity fields, and the second is that the radial peculiar velocity fluctuates on the largest scales of $\sim 0.2$ kpc$^{-1}$, whereas the rotational peculiar velocity field peaks around the smaller scale regime of $k \sim 0.5$ kpc$^{-1}$. The gas shows even more pronounced velocity fluctuations. {\bff The power spectra for the bar only model, B15, is shown to have a similar shape to those of K14, which is expected owing to the presence of a bar in the central regions of both models (although we caution that the bar in B15 is modelled only as a rotating quadrupole potential $\propto \cos 2\theta$ in the azimuthal direction). In contrast, model G13F shows little power, given the lack of non-axisymmetric structure in the centre.}

{\bff The middle panel of Fig. \ref{ps5} shows the power spectra of the peculiar velocity fields for a region that we define as the `whole disc', i.e., extending from -4.0 to 20.0 in $X$ and from -12.0 to 12.0 in $Y$. The relatively large size of the `whole disc' patch makes it possible to probe scales up to $k \sim 0.05$ kpc$^{-1}$. For our fiducial spiral model, K14, the power decreases on all scales in comparison to that of the central region, and the shape of the radial peculiar velocity power spectrum shows a single prominent peak at around scales of $k=0.2$ kpc$^{-1}$, with decreasing power on scales $k < 0.1$ kpc$^{-1}$. For the G13F spiral, the power of peculiar velocity fields is again lower than that of K14, particularly the radial velocity field, though the spectra exhibit a broad peak shape which is a consequence of the spiral structure. The power spectra of both fields are peaked at scales of around $k \sim 0.2$ kpc$^{-1}$, although the maximum power is only 5 $\rm km \; s^{-1}$, much less than the 18-20 $\rm km \; s^{-1}$ peak of K14. For B15, the distinct peak observed in the top-left panel is no longer present, while the power spectrum of the radial peculiar velocity field is reduced and distinguishable from that of K14.} 

{\bff The bottom panel of Fig. \ref{ps5} shows the power spectra for the whole disc region (as described above), in which the peculiar velocities in the central region (the same central region analysed in the top panels) are set to zero. We have re-introduced the density wave spiral models as these now contain the required data. The shape of the power spectra for all models does not appear to change from that of the whole disc upon the exclusion of the central region, and shows reduced power in most cases. The rotational peculiar velocity power spectra {\bff reveal that all simulations, except the gas component of K14, show similar features of increasing power with larger scales, and are therefore difficult to distinguish. }

{\bff The radial peculiar velocity power spectrum yields striking differences between the models presented; the barred-spiral model of K14 shows a highly peaked power spectrum that peaks on scales of $k = 0.2 \; \rm kpc ^{-1}$, whereas model G13F shows a broader peak that reaches a maximum on similar scales. In contrast, models F14 and M15 exhibit no such peak in their power spectra, with very little power on all scales. This analysis indicates that the combination of both radial and rotational peculiar velocity fields in various regions of the discs of external galaxies is required to distinguish between different spiral arm and bar models.}

\section{Conclusions}

We have presented a quantitative analysis of the scale and power of the fluctuations in peculiar velocity induced by non-axisymmetric structure, which we apply to a suite of numerical simulations. The suite comprises $N$-body simulations that naturally produce transient, winding spiral arms of varying types, and test particle simulations that treat spiral structure as an analytically prescribed density wave. We compare the results to the latest Milky Way data, and come to the following conclusions:

\begin{itemize}
\item{} From the fiducial Milky Way-type $N$-body simulation K14, we select a spiral arm that is similar to the Perseus arm of the Milky Way, and determine the solar position such that the spiral arm {\bff is $\sim 4$ kpc from the Sun in the direction $l=90$, similar to what is found by} \citet{RMB14}. {\bff We show that it is possible for the combination of a bar and transient, co-rotating spiral arms to reproduce the observed power spectrum of the APOGEE-RC and RAVE data.} However, the test particle simulations with density wave-like spirals do not fit the {\bff observed power spectrum well}, which is consistent with previous work \citep{BBG14}. {\bff However, it remains to be seen whether or not the effects of density wave spiral arms combined with a bar can reproduce the data. Therefore, analysis of test particle simulations that include a bar and spiral wave potential are a natural next step to this work, and will be presented in a forthcoming paper (Monari et al. in prep.).}
\item{} It is possible that transient, winding spiral structure alone can reproduce the defining peak feature of the observed power spectrum. This emphasises the differences between the effects of transient co-rotating spiral arms and density waves, and suggests that the power spectrum of K14 is not purely caused by the bar.
\item{} The peculiar velocity power spectrum is sensitive to the proximity of the spiral arm with respect to the solar position, the number of spiral arms (or the degree of flocculence) and the pitch angle of the spiral arm. However, further study is needed to separate out all of these dependencies fully. 
\item{} Analysis of the radial and rotational peculiar velocity fields in the central regions of discs shows that barred models can easily be separated from spiral only models.
\item{} The power spectra of whole disc regions {\bff are a more robust way of characterising the velocity fluctuations, and appear usable to distinguish between the simulations presented in this paper. Furthermore, gas motions are shown to follow those of the stars} and exhibit very similar power spectra, though with more power in general.
\end{itemize}

{\bff In this paper, we have calculated the power spectra of Fourier-transformed peculiar velocity fields primarily as a means to quantify and compare with the observational and model data of \citet{BBG14}. However, we have shown that information of both the intrinsic radial and tangential peculiar velocity field is required in order to distinguish between the different mechanisms to induce the peculiar velocity field in the Galactic disc. Therefore, it is necessary for Galactic surveys to go beyond the LOS velocity information. The full phase space information that surveys such as \emph{Gaia} will yield is expected to provide excellent constraints on dynamical models of the Galaxy and non-axisymmetric structure thereof.}

In addition to the Milky Way, we have also shown that we can make predictions for external galaxies, for some of which peculiar velocity information should be readily available from integral field spectroscopic surveys such as MUSE.

\section*{acknowledgements}
{\bff The authors thank the referee for a helpful report that lead to improvements of the manuscript. R.G. thanks Ivan Minchev for interesting discussions.} R.G. acknowledges support by the DFG Research Centre SFB-881 `The Milky Way System' through project A1. J.B. and D.K. acknowledge the generous support and hospitality of the Kavli Institute for Theoretical Physics in Santa Barbara during the `Galactic Archaeology and Precision Stellar Astrophysics' program, where some of this research was performed. This research was supported in part by the National Science Foundation under Grant No. NSF PHY11-25915. J.B. acknowledges support from a John N. Bahcall Fellowship and the W.M. Keck Foundation. GM is supported by a postdoctoral grant from the {\it Centre National d'Etudes Spatiales} (CNES). The calculations for this paper were performed on the UCL Legion, the Iridis HPC facility provided by the Centre for Innovation and the DiRAC Facilities (www.dirac.ac.uk, the DiRAC Shared Memory Processing system at the University of Cambridge operated by the COSMOS project, at the Department of Applied Mathematics and Theoretical Physics, the DiRAC Data Analytic system at the University of Cambridge, operated by the University of Cambridge High Performance Computing Service, the DiRAC Complexity system, operated by the University of Leicester IT Service, through the COSMOS consortium) jointly funded by BIS National E-infrastructure capital grant (ST/J0005673/1, ST/K001590/1 and ST/K000373/1), STFC capital grants (ST/H008861/1 and ST/H00887X/1) and STFC Operations grant (ST/K00333X/1). DiRAC is part of the UK National E-infrastructure. We also acknowledge PRACE for awarding us access to resource Cartesius based in Netherlands at SURFsara and Sisu based at CSC, Finland. This work was carried out, in part, through the Gaia Research for European Astronomy Training (GREAT-ITN) network. The research leading to these results has received funding from the European Union Seventh Framework Programme ([FP7/2007-2013]) under grant agreement number 264895.

\bibliographystyle{mn2e}
\bibliography{Vpec-R1vjun01.bbl}

\end{document}